# Energy-efficient domain wall motion governed by the interplay of helicity-dependent optical effect and spin-orbit torque

Boyu Zhang,[1,2,3] Yong Xu,[1,2] Weisheng Zhao,[1,*] Daoqian Zhu,[1] Xiaoyang Lin,[1] Michel Hehn,[2] Gregory Malinowski,[2] Dafiné Ravelosona,[3] Stéphane Mangin[2]

[1]Fert Beijing Institute, BDBC, School of Microelectronics, Beihang University, Beijing 100191, China.

[2]Institut Jean Lamour, CNRS UMR 7198, Université de Lorraine, 54000 Nancy, France.

[3]Centre de Nanoscience et de Nanotechnologie, CNRS UMR 9001, Université Paris-Sud, Université Paris-Saclay, 91400 Orsay, France.

*Corresponding author. Email: weisheng.zhao@buaa.edu.cn (W.Z.)

Spin-orbit torque provides a powerful means of manipulating domain walls along magnetic wires. However, the current density required for domain wall motion is still too high to realize low power devices. Here we experimentally demonstrate helicity-dependent domain wall motion by combining synchronized femtosecond laser pulses and short current pulses in Co/Ni/Co ultra-thin film wires with perpendicular magnetization. Domain wall can remain pinned under one laser circular helicity while depinned by the opposite circular helicity. Thanks to the all-optical helicity-dependent effect, the threshold current density due to spin-orbit torque can be reduced by more than 50%. Based on this joint effect combining spin-orbit torque and helicity-dependent laser pulses, an optoelectronic logic-in-memory device has been experimentally demonstrated. This work enables a new class of low power spintronic-photonic devices beyond the conventional approach of all-optical switching or all-current switching for data storage.

# I. INTRODUCTION

Domain wall motion is of great interest for several spintronic applications, such as high-density racetrack memory [1,2] and magnetic domain wall logic [3-5], where data can be encoded non-volatilely as a pattern of magnetic domain walls that moves along magnetic wires. The manipulation of domain walls is generally induced by external magnetic fields [6] or currents [7,8]. Current-induced domain wall motion, driven by spin-transfer torque or spin-orbit torque, has provided new opportunities for high performance all-electrical spintronics devices. However, the current density required to move domain walls is still too high [9,10] and the energy consumption still needs to be reduced.

Materials with strong perpendicular magnetic anisotropy (PMA), such as [Co/Ni] multilayers [11-13], are often considered as promising materials for current-induced magnetization reversal or domain wall motion because of their high spin polarization and tunable PMA [9-10,14-17]. In order to study the physics of domain wall motion and to provide alternative solutions for domain wall manipulation at low power consumption, several methods have been proposed, such as electric field [18] and ultrafast optics [19]. Indeed, ultrashort laser pulses provide an alternative way for magnetization manipulation [20-24]. All-optical helicity-dependent switching (AO-HDS) was observed in ferromagnetic multilayered thin films using circularly polarized light, where deterministic switching of magnetic states was achieved using laser helicity as a new degree of freedom. It has attracted significant interest because of its potential for the integration of ultra-low power all-optical writing in data storage industries [25-28].

In this work, we will show the combined effect of synchronized femtosecond laser pulses and short current pulses on domain wall motion in Co/Ni/Co ultra-thin film wires with perpendicular magnetization. The contributions from helicity-dependent optical effect, heating from laser pulses and current pulses, and spin-orbit torque were investigated by tuning the laser polarization, the current amplitude, the current pulse duration, and the synchronization delay between the electron and light stimuli. The origin of the effect has been elucidated by examining the physical contributions of different parameters using the Fatuzzo-Labrune model [29,30]. In addition, we present how such an effect can be exploited to generate Boolean logic functions. Our findings provide novel insights towards the development of low power optoelectronic logic-in-memory device, where logic functions can be implemented into memory array [31,32].

## II. RESULTS

### A. Preliminary AO-HDS and current-induced domain wall motion experiments in perpendicularly magnetized Co/Ni/Co wires

The investigated sample is a sputtered ultra-thin film of Ta(3 nm)/Pt(5 nm)/Co(0.3 nm)/Ni(0.6 nm)/Co(0.3 nm)/Pt(2 nm) on a glass substrate [Fig. 1(a)]. Thin film magnetic properties were characterized by a superconducting quantum interference device-vibration sample magnetometer (SQUID-VSM) at room temperature. We obtain a coercivity field $H_c$ of 35 Oe and a saturation magnetization $M_S$ of 770 emu/cm$^3$. The hysteresis loop measured with a magnetic field applied perpendicular to the plane of the layers shows the perpendicular anisotropy of the thin film [33].

To investigate the interplay of spin-orbit torque and helicity-dependent optical effect on the domain wall motion, microsecond current pulses and laser pulses with a pulse duration of 35 femtoseconds (fs) were synchronized at a 5 kHz repetition rate [Fig. 1(a)]. A Kerr microscope was used to image the magnetic configuration of the investigated samples. The magnetic contrast was enhanced by subtracting two pictures taken before and after the domain wall nucleation, current injection or laser excitation, where the initial magnetization saturation direction is up (M↑) and the white contrast corresponds to a reversal to down magnetization (M↓).

First, the AO-HDS process and current-induced domain wall motion process were measured independently. After saturating the thin films under a perpendicular magnetic field, the laser beam was swept over the film surface without any applied magnetic field or current. A clear AO-HDS effect was observed in the Co/Ni/Co ultra-thin film as reported in Fig. 1(b) for a fluence of

9 mJ/cm$^2$. The sample was then patterned into 4 μm width wires by UV optical lithography and ion beam etching. After saturation under a perpendicular magnetic field, the laser beam with a fluence of 9 mJ/cm$^2$ was swept along the magnetic wire under no magnetic field or current. Fig. 1(c) indicates that the AO-HDS process is still observed in the 4 μm Co/Ni/Co ferromagnetic wire after the patterning. Furthermore, current pulses with a current density of 16.5×10$^6$ A/cm$^2$, a pulse duration of 8 μs and a frequency of 5 kHz were applied during 10 seconds without any magnetic field or laser beam after the nucleation of a reversed domain in the wire. This results in domain wall propagation, as shown in Fig. 1(d). Indeed, the presence of the Pt layer can induce spin Hall effect (SHE), which converts the charge currents into pure spin currents perpendicular to the electrical current [34]. In our samples, both Pt layers act as a spin current source with opposite spin direction. However, since the spin Hall current depends on the layer thickness [35], a net spin Hall current is expected to occur in our sample with a direction of the effective field $H_{SHE}$ that is along $\boldsymbol{m}\times(\boldsymbol{z}\times\boldsymbol{j_e})$ [36], where $\boldsymbol{m}$, $\boldsymbol{z}$ and $\boldsymbol{j_e}$ are unit vectors along the magnetization, z axis and electron flow, respectively. With an effective Dzyaloshinskii–Moriya interaction (DMI) field $H_{DMI}$ of +300 Oe [33] that favors a left-handed chiral Néel domain walls [37,38], $H_{SHE}$ enables the domain wall propagation against the electron flow, which is consistent in our case with current-induced domain wall motion driven by spin-orbit torque. After the nucleation of a reversed domain in the wire, a laser fluence lower than 4.5 mJ/cm$^2$ with a 5 kHz repetition rate and a 35 fs pulse duration is not sufficient to move the domain wall. Also, a current density lower than 15×10$^{16}$ A/cm$^2$ with a 5 kHz frequency and a 10 μs pulse duration cannot move the domain wall. Those quantities define the laser fluence and the current density threshold values.

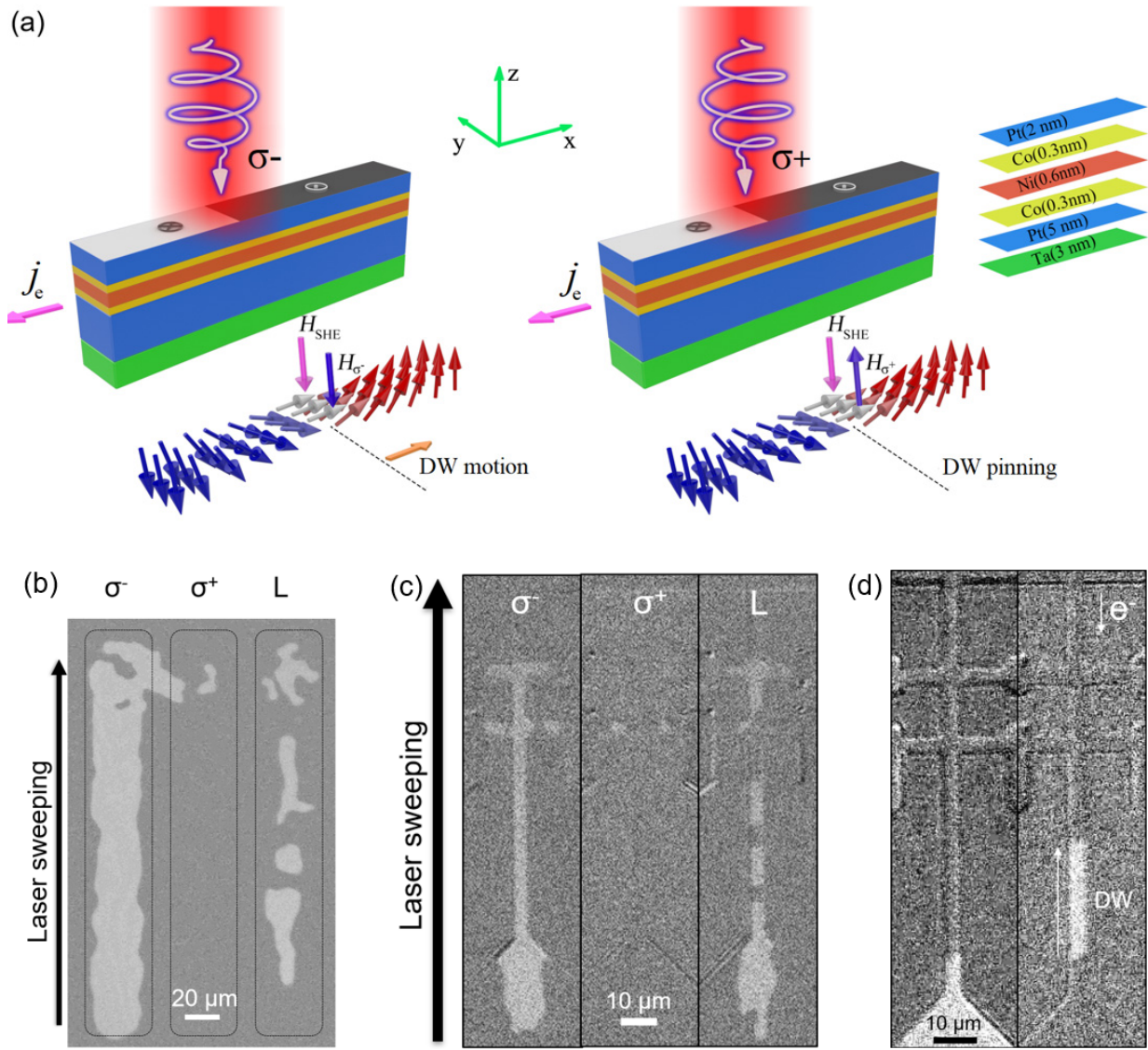

FIG. 1. (a) Schematic of the experimental set-up: A current pulse is injected along the 4 μm Co/Ni/Co ferromagnetic wire while a synchronized femtosecond laser beam shines on the wire. Grayscale topside displays Kerr images from a left-handed chiral Néel domain wall nucleated in the wire. The effective field $H_{SHE}$ moves domain walls in the direction against electron flow $j_e$, while the right-circularly (σ+) or left-circularly (σ-) polarized laser beam favors M↑ or M↓. (b) Kerr images of the Co/Ni/Co thin film. Linear (L), right-circularly (σ+) and left-circularly (σ-) polarized laser beam were swept over the film with a fluence of 9 mJ/cm$^2$. (c-d) Kerr images of 4 μm Co/Ni/Co ferromagnetic wires. (c) Linear (L), right-circularly (σ+) and left-circularly (σ-) polarized laser beam were swept over the wire with a fluence of 9 mJ/cm$^2$. (d) Current pulses

with a density of 16.5×10$^6$ A/cm$^2$, a pulse duration of 8 μs and a frequency of 5 kHz were injected during 10 seconds after a nucleation of a reversed domain in the wire. The initial magnetization saturation direction is M↑ and the white contrast corresponds to a reversal to magnetization M↓.

## B. Domain wall motion combining synchronized femtosecond laser pulses and short current pulses

Based on the presence of AO-HDS and current-induced domain wall motion in 4 μm Co/Ni/Co ferromagnetic wires, the experiments combining helicity-dependent optical effect and spin-orbit torque were performed with current densities and laser fluences below threshold values defined in the previous section, and a single domain wall was priorly nucleated in the wire after a magnetization saturation with M↑.

Fig. 2(a) shows domain wall motion after 10 seconds of laser pulses with 4 mJ/cm$^2$ fluence and 5 kHz repetition rate together with synchronized current pulses of 7.3×10$^6$ A/cm$^2$ and 10 μs pulse duration. The center of the laser spot is placed 8 μm away from the domain wall, as indicated by the star in Fig. 2(a). The left-circularly (σ-) polarized laser beam induces a larger domain wall displacement since the domain wall moves further than the position of the beam, while linear (L) polarized laser beam induces a moderate domain wall propagation up to the beam center, whereas the right-circularly (σ+) polarized laser beam doesn't lead to any domain wall motion. The beam center corresponds to the hottest region [inset of Fig. 2(a)], which induces the domain wall propagation up to this point with linear (L) due to Gaussian laser heating. The measurements obtained for the two helicities demonstrate that domain wall can remain pinned when the laser beam shines with one circular helicity while it is depinned when using the opposite circular helicity. This clearly demonstrates that the laser polarization offers a new degree of freedom to control domain wall motion. It also shows that the domain wall moves against the direction of the electron flow, which confirms the effect of spin-orbit torque. The dashed lines in Fig. 2 correspond to the initial domain wall position. With the reversed current, the domain wall moves in the opposite direction as a propagation towards M↓ domains, which is also against the direction of electron flow, and the role of left-circularly (σ-) and right-circularly (σ+) polarized laser reverses as σ+ favors a propagation towards M↓ domains [33]. With the inputs of synchronized laser and current pulses below threshold, a large domain wall displacement can only be realized with both left-circularly (σ-) polarized laser beam and current injection, which can be exploited for magnetic domain wall logic.

The domain wall motion was further studied by increasing the current density up to $14.6 \times 10^6$ A/cm$^2$, which is still below the threshold value for domain wall motion. The center of the laser spot was placed 20 μm away from the domain wall, as indicated by the star in the inset of Fig. 2(b). Because of the Gaussian distribution of the laser intensity, the domain wall can propagate over a large distance as shown in the inset of Fig. 2(b), as a linear (L) laser beam shines on the sample, where the dashed lines correspond to the initial domain wall position. The video of the domain wall motion was recorded and the time evolution of the domain wall velocity was obtained by detecting the distance travelled by the domain wall along the wire using the APREX TRACK software [39]. By the integration of the domain wall velocity in regard to the time evolution, the displacement as a function of the laser polarization is shown in Fig. 2(b). Three regimes can be distinguished: domain wall propagates slowly at first and moves rapidly as it gets close to the center of the laser spot, then it slows down and stops. The contribution of Gaussian laser heating around the beam spot on the domain wall motion is confirmed. With the increased current density, spin-orbit torque enables the domain wall propagation regardless of the laser polarization. Still some helicity-dependence optical effect is clearly demonstrated and left-circularly (σ-) polarized laser beam gives the maximum velocity for a domain wall propagation towards M↑ domains. With the reversed current, the domain wall moves in the opposite direction as a propagation towards M↓ domains, which is also against the direction of electron flow, and right-circularly (σ+) polarized laser beam induces a larger velocity as it favors a propagation towards M↓ domains [33]. The experiments were repeated, and the above conclusions can be confirmed.

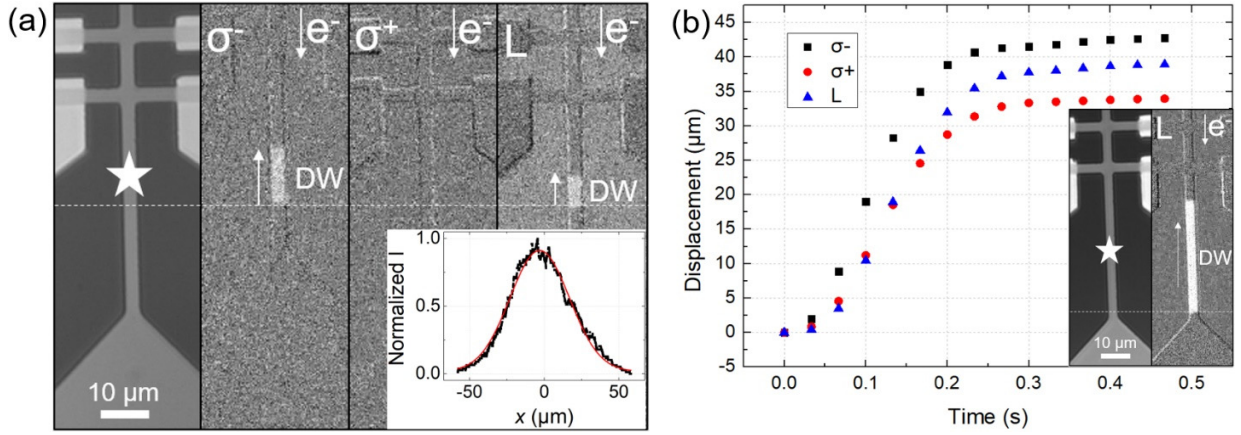

FIG. 2. (a) Kerr images of a 4 μm Co/Ni/Co ferromagnetic wire. Linear (L), right-circularly (σ+) and left-circularly (σ-) polarized laser beams shine on the sample with a fluence of 4 mJ/cm$^2$ for 10 seconds, together with synchronized current pulses of $7.3\times10^6$ A/cm$^2$ and 10 μs pulse duration (5 kHz repetition rate), both of which are below the threshold value for domain wall motion. Inset: Gaussian profile of the laser spot with a full width at half maximum (FWHM) of 47 μm. (b) Time evolution of the domain wall displacement of a 4 μm Co/Ni/Co ferromagnetic wire from domain wall video recording and analysis with the APREX TRACK software as a function of the laser polarization, where the current density was increased to $14.6\times10^6$ A/cm$^2$. Inset: Kerr images of a 4 μm Co/Ni/Co ferromagnetic wire, which shows the domain wall motion with Linear (L) laser beam. The initial magnetization saturation direction is M↑ and the white contrast corresponds to a reversal to M↓.

**C. Effect of synchronization delay between the electron and light stimuli on domain wall motion**

In order to determine the respective contributions of current pulses and laser pulses, the synchronization delay between the electron and light stimuli, and the duration of current pulses were varied. In this experiment, the center of the laser spot was placed close to the domain wall, as indicated by the star in Fig. 3(a). During 10 seconds, linear (L) laser pulses with a 4 mJ/cm$^2$ fluence and current pulses of 11×10$^6$ A/cm$^2$ were injected in the wire with a 5 kHz repetition rate. The synchronization delay between the laser pulses and the current pulses was then adjusted. Both the laser fluence and the current density were below the threshold values for domain wall motion, and a single domain wall was priorly nucleated in the wire after a magnetization saturation with M↑.

Results are shown in Fig. 3(a). For both 5 μs and 10 μs current pulse durations, we can observe that the domain wall propagates the same distance if the laser pulses and the current pulses are synchronized. In this case, the two effects act simultaneously and the domain wall propagation is limited by the competition between the Gaussian laser heating, spin-orbit torque and the pinning. The bottom dashed line corresponds to the initial domain wall position. The study on the effect of the synchronization delay on the domain wall motion shows that the domain wall propagates less when the current pulse is injected after the laser pulse and the domain wall propagation distance is smaller for 100 μs delay than 50 or 150 μs. As the absorption of laser energy occurs only in the electronic bath, the laser pulses induce a large increase of electron temperature $T_e$. Then, the electronic thermal bath is coupled to that of the phonons, which induces a fast decrease of temperature to a stable value. The center of the laser beam spot normally corresponds to a maximum temperature of 600 K [19]. The temperature increases $\Delta T$ due to the current heating from an injected current pulse [40] are 38 K or 59 K for the pulse duration of 5 μs or 10 μs, respectively [33]. The wire temperature increase as a function of time can be estimated, as shown in Fig. 3(b). When the current pulse comes 50μs after the laser pulse, the residual laser heating helps the spin-orbit torque to move the domain wall. When the delay is 100 μs, the residual laser heating is small when the following current pulse is injected, while the residual heating of this current pulse is small when the following laser pulse arrives, but they can induce a shorter domain wall displacement distance. In the case of 150 μs, a larger displacement can be obtained due to the larger residual current heating. The table of Fig. 3(a) shows the domain wall displacement distance for each pulse duration and synchronization delay. The integration of residual temperature increases $\Delta T$ in regard to the time is indicated by the blue shadow area in Fig. 3(b) for each pulse duration and synchronization delay, where a larger blue shadow area (integration of $\Delta T$ in regard to the time) induces a larger longer domain wall displacement. The larger displacement with 10 μs current pulse duration than that with 5 μs shows the role of current heating, as 10 μs current pulse induces a larger $\Delta T$ with a longer applied time.

The contribution of current heating and spin-orbit torque on the domain wall motion is demonstrated by performing field-driven current-assisted domain wall velocity measurements [33], indicating that the current heating increases the domain wall velocity by helping the domain wall to overcome the pinning energy barrier [41], while spin-orbit torque assists the domain wall motion against the direction of electron flow.

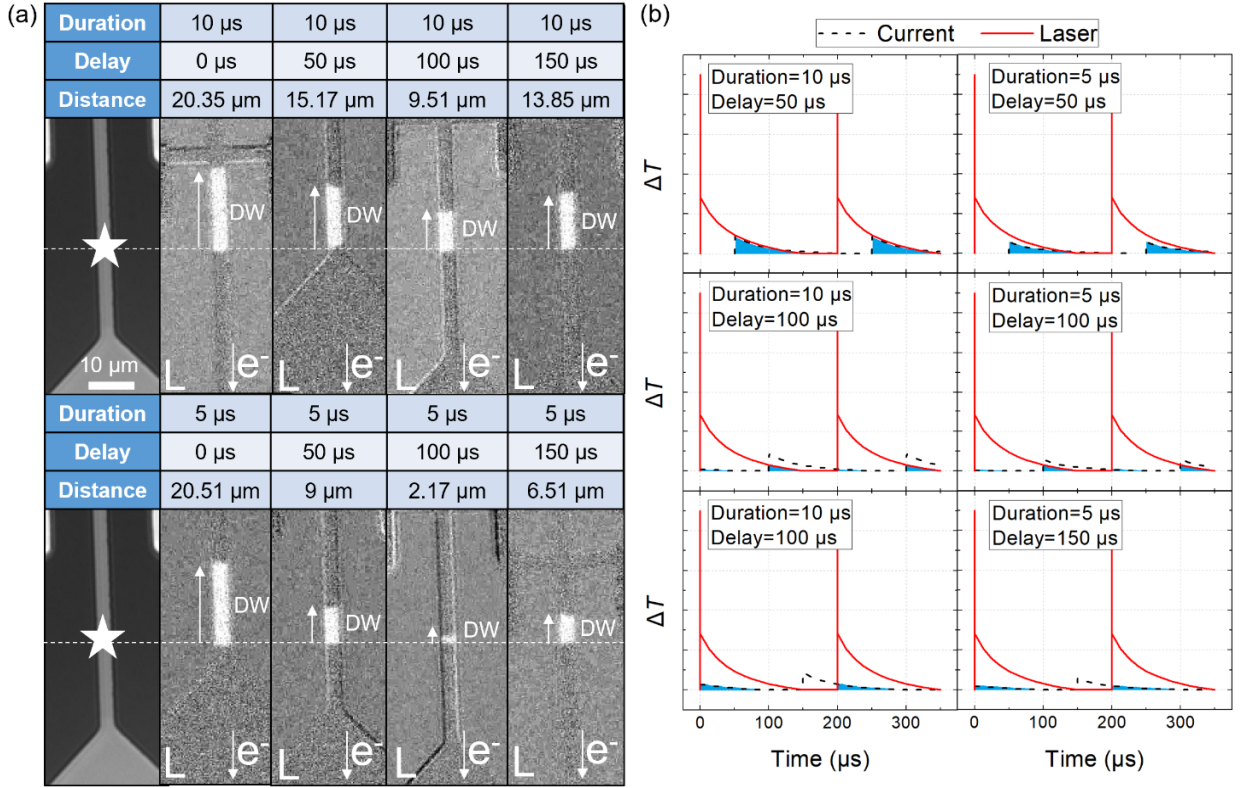

FIG. 3. (a) Kerr images of 4 μm Co/Ni/Co ferromagnetic wires. Linear (L) laser beam shines on the sample with a fluence of 4 mJ/cm$^2$ for 10 seconds, together with current pulses of 11×10$^6$ A/cm$^2$ and 10 μs or 5μs pulse duration with a synchronization delay of 0, 50, 100, 150 μs (200 μs period), both of which are below the threshold values for domain wall motion. The initial magnetization saturation direction is M↑ and the white contrast corresponds to a reversal to M↓. The domain wall displacement distances are shown in the table for each pulse duration and synchronization delay. (b) Schematic of the wire temperature increase Δ$T$ due to the laser pulses (black solid lines) and the current pulses (red dashed lines) as a function of time with the synchronization delay of 50, 100 or 150 μs (200 μs period). The integration of residual temperature increases Δ$T$ in regard to the time is indicated as a blue shadow area for each pulse duration and synchronization delay.

**D. Energy consumption**

To quantify the energy consumption in the presence of laser pulses, we have compared the current-induced domain wall motion as shown in Fig. 1(d) and the domain wall motion combining synchronized current pulses and laser pulses as shown in Fig. 2(a) with left-circularly (σ-) polarized laser beam. With a total energy of 14.6×10$^{-4}$ J per 1 μm domain wall displacement with current density of 16.5×10$^6$ A/cm$^2$ for the 1$^{st}$ case and 6.5×10$^{-4}$ J/μm with 7.3×10$^6$ A/cm$^2$ for the 2$^{nd}$ case [33], we can conclude that in the presence of circularly polarized laser pulses, the energy consumption and the threshold current density due to spin-orbit torque for domain wall motion can be reduced by more than 50% in the investigated wire.

## III. DISCUSSION

The above results demonstrate that domain wall motion is due to the combination of helicity-dependent optical effect, heating from laser pulses and current pulses, and spin-orbit torque. In order to analyze the different contributions, we have used the Fatuzzo-Labrune model [29,30] to evaluate the domain wall velocity and displacement under the combined action of laser pulses and current pulses:

$$v = v_0 \exp(-\frac{E - 2H_{eff}M_S V_B}{k_B T}) \qquad (1)$$

where $E$ represents the pinning energy barrier to be overcome in order to enable the domain wall motion within the Barkhausen volume $V_B$, $H_{eff}$ is an effective field that contains the contribution from the laser pulses and the current pulses.

The spin-orbit torque can be described by an effective field $H_{SHE}$ originating from the SHE whose direction favors domain wall propagation against the direction of electron flow [9]: $H_{SHE}=(\hbar\theta_{SH}J)/(2eM_S t)$, where $\hbar=h/2\pi$ and $h$ is the Planck constant, $e$ represents the elementary charge, $\theta_{SH}$ is the Spin Hall angle, $t$ stands for the total thickness of the ferromagnetic layers. The Gaussian laser heating due to the laser profile [inset of Fig. 2(a)] gives a Gaussian distribution of temperature $T$ and we assume that the center of the laser beam corresponds to a maximum temperature of 600 K for linear polarization [19]. The laser helicity induces an effective field $H_\sigma$ and the direction of $H_\sigma$ depends on the laser helicity.

A current density $J$ of $7.3\times10^6$ A/cm$^2$ gives a $H_{SHE}$ of 16 Oe with $M_S$=770 emu/cm$^3$, $t$=1.2 nm and $\theta_{SH}$=0.04, and also a temperature increase $\Delta T$ of 25.8 K to $T$ [33]. σ- and σ+ induce the $H_\sigma$ that favors M↓ and M↑, respectively, where $|H_\sigma|$ equals to 3 Oe. As the domain wall velocity $v$ is a function of the Gaussian distribution of $H_\sigma$ and $T$ related to the laser position $x$ with $v=dx/dt$ $=f(x)$, by solving the equation, the domain wall displacement $x$ as a function of the time $t$ can then be obtained with $V_B$=10$^{-23}$ m$^{-3}$, $v_0$=2×10$^{12}$ μm/s, as shown in Fig. 4(a). Spin-orbit torque enables the domain wall motion against the direction of the electron flow. Domain wall motion moves significantly with left-circularly (σ-) laser pulses due to the laser helicity, and Gaussian laser heating enables the domain wall motion to the beam center with linear (L) light, while right-circularly (σ+) laser pulses give almost no domain wall motion. The simulations as shown in Fig. 4(a) tend to indicate the dominated role of the laser polarization with smaller current density, which provide the domain wall motion results similar to those previously described in Fig. 2(a).

With a larger $H_{SHE}$ of 32 Oe and a higher $\Delta T$ of 103.3 K [33] corresponding to $J$ of 14.6×10$^6$ A/cm$^2$, the domain wall displacement is shown in Fig. 4(b), where the larger spin-orbit torque induces the domain wall motion regardless of the laser polarization. The three regimes of domain wall displacement confirm the effect of Gaussian laser heating, while SOT plays a dominant role on the domain wall velocity, which is in agreement with the experimental results as shown in Fig. 2(b).

In addition, based on the above conditions, implementing only current pulses or laser pulses into the model gives a vanishing domain wall velocity and displacement profile. Therefore, the main results of domain wall motion experiments combining current pulses and laser pulses are well reproduced.

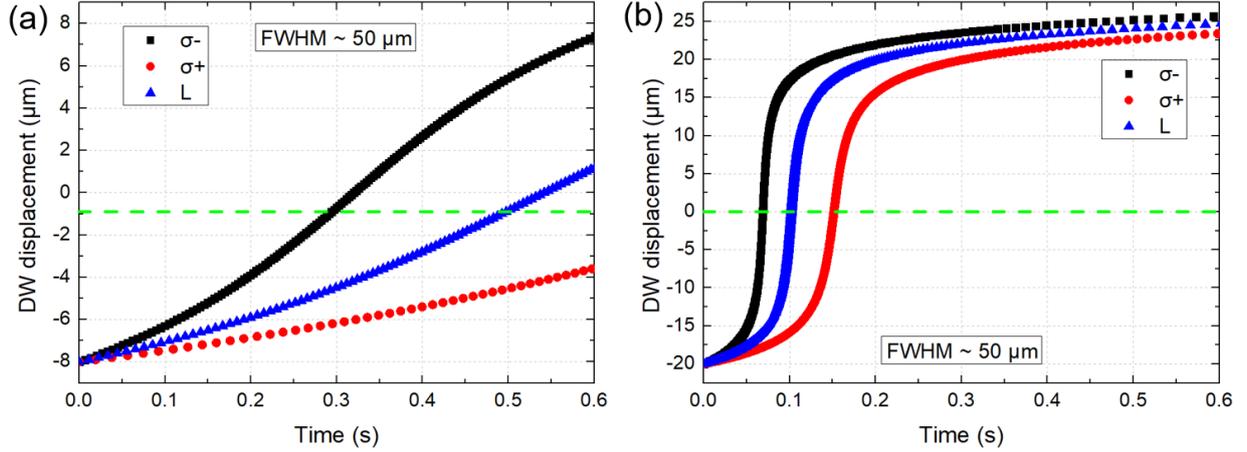

FIG. 4. Time-dependent simulations of the domain wall displacement based on the Fatuzzo-Labrune model. The domain wall motion is induced by synchronized current pulses and linear (L), right-circularly (σ+) or left-circularly (σ-) polarized laser pulses with (a) a $H_{SHE}$ of 12 Oe, a Gaussian distribution of $H_σ$ and $T$ with a FWHM of ~50 μm, (b) a $H_{SHE}$ of 24 Oe, a Gaussian distribution of $H_σ$ and $T$ with a FWHM of ~50 μm. Similar results for the domain wall displacement as shown in Fig. 2(a) and Fig. 2(b) are obtained in (a) and (b), respectively. The green dashed line corresponds to the position of the laser beam center.

Based on the domain wall motion resulting from the contribution of the helicity-dependent optical effect and the spin-orbit torque, we propose that our structure, which mimics a magnetic domain wall gate, can form the building block for generating Boolean logic functions. An example for an AND gate is given in Fig. 5. The principle is very similar to the concept of the magnetic shift register based on moving domain walls in a racetrack geometry [1-5]. The elementary logic device consists of a domain wall wire, with two inputs 'A' and 'B' that involve synchronized injected current and shined left-circularly (σ-) polarized laser beam, both of which are below the threshold values for domain wall motion, and an output 'C' that involves detecting the local magnetization through the anomalous Hall effect (AHE). The device functions as follows: First, a single domain wall is nucleated in the wire, away from the AHE detection area, with the output 'C' set to the resistance state '0' (M↑). If none or either of the stimuli 'A' and 'B' are applied, no domain wall motion is induced, which leaves the output 'C' in the '0' state. However, if both stimuli are on (inputs 'A' and 'B' set to '1'), the generated domain wall can propagate along the wire, driven by the combination of spin-orbit torque 'A' and the helicity-dependent optical effect 'B', leading to the resistance state '1' (M↓) of the output 'C'. Therefore, a promising approach towards low power spintronic-photonic logic device can be constructed using domain wall motion in magnetic wires.

In addition, after nucleation of a single domain wall and set of output 'C' in the resistance state '0' (M↑), implementing left-circularly (σ-) or right-circularly (σ+) polarized laser beam for input 'A' along with current for input 'B' leads to the '1' or '0' state for output 'C', which can be used for direct detection of laser helicity.

As the domain walls store data non-volatilely into magnetic states, the proposed domain wall logic can be used for the logic-in-memory applications, where nonvolatile memory elements are distributed over a logic-circuit plane [31,32].

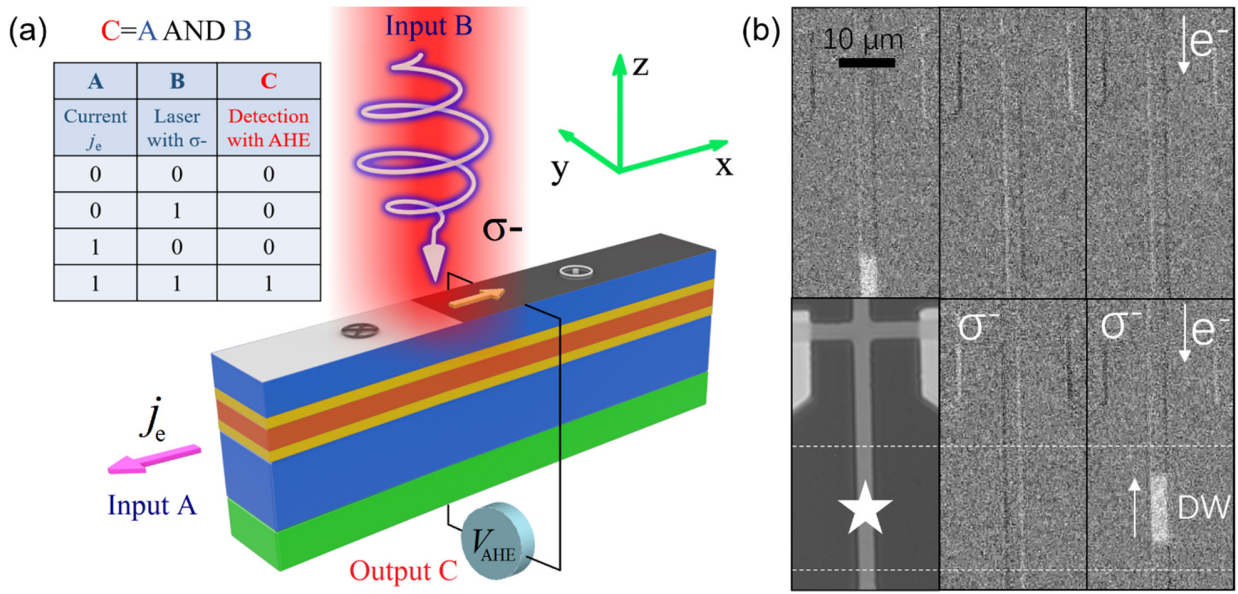

FIG. 5. (a) Design of an AND logic function by using laser pulses and current pulses to control the domain wall motion. Inputs 'A' and 'B' are synchronized current pulses and laser pulses below the threshold values for domain wall motion, and output 'C' serves as read out for the magnetization direction in the wire through AHE. This scheme corresponds to a logical AND operation and AND table is shown on top. (b) Kerr images of 4 μm Co/Ni/Co ferromagnetic wires. The current pulses and laser pulses are the same with Fig. 2(a). The top three figures correspond to the domain wall nucleation, domain wall motion with input (A,B)=(0,0) and (1,0), while the bottom three figures correspond to the indication of the laser spot center in the wire as a star, domain wall motion with (A,B)=(0,1) and (1,1). The initial magnetization saturation direction is M↑ and the white contrast corresponds to a reversal to M↓.

## IV. CONCLUSION

In summary, we have experimentally demonstrated helicity-dependent domain wall motion by the combined effect of synchronized femtosecond laser pulses and short current pulses in Co/Ni/Co ultra-thin film wires with perpendicular magnetization. The domain wall motion results from the interplay of helicity-dependence optical effect, heating from laser pulses and current pulses, and spin-orbit torque. The laser polarization provides a new degree of freedom to manipulate the domain wall in magnetic devices, where domain wall can remain pinned under one laser circular helicity while depinned for the opposite circular helicity. Due to the contribution of helicity-dependence optical effect, the energy consumption and the threshold current density due to spin-orbit torque for domain wall motion can be reduced by more than 50% in the investigated wire. Our energy-efficient approach highlights a new path towards low power optoelectronic logic-in-memory devices, which enables the development of new families of spintronic devices combining photonics and electronics.

## ACKNOWLEDGMENTS

The authors thank Prof. Eric Fullerton and Prof. Nicolas Vernier for fruitful discussions, and Dr. Laurent Badie for technical assistance with the lithography process. The authors gratefully acknowledge the National Natural Science Foundation of China (Grant No. 61627813, 61571023, 51602013), Program of Introducing Talents of Discipline to Universities (No. B16001) , the National Key Technology Program of China (No. 2017ZX01032101), Young Elite Scientists

Sponsorship Program by CAST (No. 2018QNRC001) and the China Scholarship Council. This work was supported by the ANR-15-CE24-0009 UMAMI and by the ANR-Labcom Project LSTNM, by the Institut Carnot ICEEL for the project « Optic-switch » and Matelas, and by the French PIA project 'Lorraine Université d'Excellence', reference ANR-15-IDEX-04-LUE. Experiments were performed using equipment from the TUBE. Davm funded by FEDER (EU), ANR, Région Grand Est and Metropole Grand Nancy.

**REFERENCES**

[1] S. Parkin and S.-H. Yang, Memory on the racetrack, *Nat. Nanotechnol.* **10**, 195–198 (2015).

[2] S. Parkin, M. Hayashi, and L. Thomas, Magnetic Domain-Wall Racetrack Memory, *Science* **320**, 190–194 (2008).

[3] D. A. Allwood, G. Xiong, M. D. Cooke, C. C. Faulkner, D. Atkinson, N. Vernier, and R. P. Cowburn, Submicrometer ferromagnetic NOT gate and shift register, *Science* **296**, 2003–2006 (2002).

[4] D. A. Allwood, G. Xiong, C. C. Faulkner, D. Atkinson, D. Petit, and R. P. Cowburn, Magnetic domain-wall logic, *Science* **309**, 1688-1692 (2005).

[5] J. H. Franken, H. J. M. Swagten, and B. Koopmans, Shift registers based on magnetic domain wall ratchets with perpendicular anisotropy, *Nat. Nanotechnol.* **7**, 499–503 (2012).

[6] P. J. Metaxas, J. P. Jamet, A. Mougin, M. Cormier, J. Ferré, V. Baltz, B. Rodmacq, B. Dieny, and R. L. Stamps, Creep and flow regimes of magnetic domain-wall motion in ultrathin Pt/Co/Pt films with perpendicular anisotropy, *Phys. Rev. Lett.* **99**, 217208 (2007).

[7] D. Ravelosona, S. Mangin, Y. Lemaho, J. A. Katine, B. D. Terris, and E. E. Fullerton, Domain wall creation in nanostructures driven by a spin-polarized current, *Phys. Rev. Lett.* **96**, 186604 (2006).

[8] C. Burrowes, D. Ravelosona, C. Chappert, S. Mangin, E. E. Fullerton, J. A. Katine, and B. D. Terris, Role of pinning in current driven domain wall motion in wires with perpendicular anisotropy, *Appl. Phys. Lett*. **93**, 172513 (2008).

[9] K.-S. Ryu, L. Thomas, S.-H. Yang, and S. Parkin, Chiral spin torque at magnetic domain walls, *Nat. Nanotechnol*. **8**, 527–533 (2013).

[10] S.-H. Yang, K.-S. Ryu, and S. Parkin, Domain-wall velocities of up to 750 m s$^{-1}$ driven by exchange-coupling torque in synthetic antiferromagnets, *Nat. Nanotechnol*. **10**, 221–226 (2015).

[11] G. H. O. Daalderop, P. J. Kelly, and F. J. A. Den Broeder, Prediction and confirmation of perpendicular magnetic anisotropy in Co/Ni multilayers, *Phys. Rev. Lett*. **68**, 682–685 (1992).

[12] S. Girod, M. Gottwald, S. Andrieu, S. Mangin, J. Mccord, E. E. Fullerton, J. L. Beaujour, and A. D. Kent, Strong perpendicular magnetic anisotropy in Ni/Co(111) single crystal superlattices, *Appl. Phys. Lett*. **94**, 262504 (2009).

[13] S. Andrieu, T. Hauet, M. Gottwald, A. Rajanikanth, L. Calmels, A. M. Bataille, F. Montaigne, S. Mangin, E. Otero, P. Ohresser, P. Le Fèvre, F. Bertran, A. Resta, A. Vlad, A. Coati, and Y. Garreau, Co/Ni multilayers for spintronics: High spin polarization and tunable magnetic anisotropy, *Phys. Rev. Mater*. **2**, 064410 (2018).

[14] D. Ravelosona, S. Mangin, J. A. Katine, E. E. Fullerton, and B. D. Terris, Threshold currents to move domain walls in films with perpendicular anisotropy, *Appl. Phys. Lett*. **90**, 2007–2009 (2007).

[15] C. Burrowes, A. P. Mihai, D. Ravelosona, J. Kim, C. Chappert, L. Vila, A. Marty, Y. Samson, I. Tudosa, E. E. Fullerton, and J. Attané, Non-adiabatic spin-torques in narrow magnetic domain walls, *Nat. Phys*. **6**, 17–21 (2009).

[16] T. Koyama, D. Chiba, K. Ueda, K. Kondou, H. Tanigawa, S. Fukami, T. Suzuki, N. Ohshima, N. Ishiwata, Y. Nakatani, K. Kobayashi, and T. Ono, Observation of the intrinsic pinning of a magnetic domain wall in a ferromagnetic nanowire, *Nat. Mater.* **10**, 194–197 (2011).

[17] S. Le Gall, N. Vernier, F. Montaigne, A. Thiaville, J. Sampaio, D. Ravelosona, S. Mangin, S. Andrieu, and T. Hauet, Thermally activated domain wall motion in [Co/Ni](111) superlattices with perpendicular magnetic anisotropy, *Appl. Phys. Lett.* **106**, 062406 (2015).

[18] T. H. E. Lahtinen, K. J. A. Franke, and S. van Dijken, Electric-field control of magnetic domain wall motion and local magnetization reversal, *Sci. Rep.* **2**, 258 (2012).

[19] Y. Quessab, R. Medapalli, M. S. El Hadri, M. Hehn, G. Malinowski, E. E. Fullerton, and S. Mangin, Helicity-dependent all-optical domain wall motion in ferromagnetic thin films, *Phys. Rev. B* **97**, 054419 (2018).

[20] E. Beaurepaire, J. C. Merle, A. Daunois, and J. Y. Bigot, Ultrafast spin dynamics in ferromagnetic nickel, *Phys. Rev. Lett.* **76**, 4250–4253 (1996).

[21] C. D. Stanciu, F. Hansteen, A. V. Kimel, A. Kirilyuk, A. Tsukamoto, A. Itoh, and T. Rasing, All-optical magnetic recording with circularly polarized light, *Phys. Rev. Lett.* **99**, 047601 (2007).

[22] A. Kirilyuk, A. V. Kimel, and T. Rasing, Ultrafast optical manipulation of magnetic order, *Rev. Mod. Phys.* **82**, 2731–2784 (2010).

[23] M. S. El Hadri, M. Hehn, G. Malinowski, and S. Mangin, Materials and devices for all-optical helicity-dependent switching, *J. Phys. D. Appl. Phys*. **50**, 133002 (2017).

[24] P. Liu, X. Lin, Y. Xu, B. Zhang, Z. Si, K. Cao, J. Wei , and W. Zhao, Optically tunable magnetoresistance effect: From mechanism to novel device application, *Materials*. **11**, 47 (2017).

[25] S. Mangin, M. Gottwald, C. H. Lambert, D. Steil, V. Uhlíř, L. Pang, M. Hehn, S. Alebrand, M. Cinchetti, G. Malinowski, Y. Fainman, M. Aeschlimann, and E. E. Fullerton, Engineered materials for all-optical helicity-dependent magnetic switching, *Nat. Mater.* **13**, 286–292 (2014).

[26] C. H. Lambert, S. Mangin, B. S. D. C. S. Varaprasad, Y. K. Takahashi, M. Hehn, M. Cinchetti, G. Malinowski, K. Hono, Y. Fainman, M. Aeschlimann, and E. E. Fullerton, All-optical control of ferromagnetic thin films and nanostructures, *Science* **345**, 1337–1340 (2014).

[27] M. S. El Hadri, P. Pirro, C.-H. Lambert, S. Petit-Watelot, Y. Quessab, M. Hehn, F. Montaigne, G. Malinowski, and S. Mangin, Two types of all-optical magnetization switching mechanisms using femtosecond laser pulses, *Phys. Rev. B* **94**, 064412 (2016).

[28] M. S. El Hadri, M. Hehn, P. Pirro, C. H. Lambert, G. Malinowski, E. E. Fullerton, and S. Mangin, Domain size criterion for the observation of all-optical helicity-dependent switching in magnetic thin films, *Phys. Rev. B* **94**, 064419 (2016).

[29] E. Fatuzzo, Theoretical considerations on the switching transient in ferroelectrics, *Phys. Rev.* **127**, 1999–2005 (1962).

[30] M. Labrune, S. Andrieu, F. Rio, and P. Bernstein, Time dependence of the magnetization process of RE-TM alloys, *J. Magn. Magn. Mater.* **80**, 211–218 (1989).

[31] H. S. Stone, A Logic-in-Memory Computer, *IEEE Trans. Comput.* **C-19**, 73–78 (1970).

[32] S. Matsunaga, J. Hayakawa, S. Ikeda, K. Miura, H. Hasegawa, T. Endoh, H. Ohno, and T. Hanyu, Fabrication of a nonvolatile full adder based on logic-in-memory architecture using magnetic tunnel junctions, *Appl. Phys. Express* **1**, 0913011–0913013 (2008).

[33] See Supplemental Material at [URL will be inserted by publisher] for more detailed information about the experiments, the hysteresis loop of the thin film, the DMI measurement, the domain wall motion combining laser pulses and current pulses with

reversed current and magnetization, the calculation of temperature increases due to the current heating, the role of current pulses on domain wall motion and the calculation of energy consumption.

[34] A. Hoffmann, Spin Hall Effects in Metals, *IEEE Trans. Magn.* **49**, 5172–5193 (2013).

[35] L. Liu, T. Moriyama, D. C. Ralph, and R. A. Buhrman, Spin-torque ferromagnetic resonance induced by the spin Hall effect, *Phys. Rev. Lett.* **106**, 036601 (2011).

[36] S. Emori, U. Bauer, S. M. Ahn, E. Martinez, and G. S. D. Beach, Current-driven dynamics of chiral ferromagnetic domain walls, *Nat. Mater.* **12**, 611–616 (2013).

[37] S. Je, D. Kim, S. Yoo, B. Min, K. Lee, and S. Choe, Asymmetric magnetic domain-wall motion by the Dzyaloshinskii-Moriya interaction, *Phys. Rev. B* **88**, 214401 (2013).

[38] J.-P. Tetienne, T. Hingant, L. J. Martínez, S. Rohart, A. Thiaville, L. H. Diez, K. Garcia, J.-P. Adam, J.-V. Kim, J.-F. Roch, I. M. Miron, G. Gaudin, L. Vila, B. Ocker, D. Ravelosona, and V. Jacques, The nature of domain walls in ultrathin ferromagnets revealed by scanning nanomagnetometry, *Nat. Commun.* **6**, 6733 (2015).

[39] https://www.aprex-solutions.com/

[40] J. Curiale, A. Lemàtre, G. Faini, and V. Jeudy, Track heating study for current-induced domain wall motion experiments, *Appl. Phys. Lett.* **97**, 243505 (2010).

[41] O. Boulle, G. Malinowski, and M. Kläui, Current-induced domain wall motion in nanoscale ferromagnetic elements, *Mater. Sci. Eng. R Reports* **72**, 159–187 (2011).